\documentclass[11pt,a4paper]{article}
\pdfoutput=1
\usepackage{jheppub}

\usepackage{multirow, graphicx,amssymb,url,mathrsfs,amsmath}
\usepackage{wrapfig,boxedminipage,setspace,subfigure,epsfig}
\usepackage{amsxtra,amstext,latexsym,dsfont,amsfonts}
\usepackage{color}
\usepackage[dvipsnames]{xcolor}
\usepackage{float}
\usepackage{slashed,comment}
\usepackage{contour,xparse}
\usepackage{tikz}
\usetikzlibrary{calc,patterns,angles,quotes}
\usetikzlibrary{decorations.pathreplacing,decorations.markings,snakes}
\usepackage{calligra}
\DeclareFontShape{T1}{calligra}{m}{n}{<->s*[2.2]callig15}{}
\DeclareMathAlphabet{\mathcalligra}{T1}{calligra}{m}{n}

\usepackage{tabularx, array}
\newcolumntype{L}[1]{>{\raggedright\arraybackslash}p{#1}}
\newcolumntype{C}[1]{>{\centering\arraybackslash}p{#1}}
\newcolumntype{R}[1]{>{\raggedleft\arraybackslash}p{#1}}

\newcommand{\dd}{\mathrm{d}}

\newcommand{\nr}{{\mathcalligra{r}}}

\title{Brickwall model for hyperbolic black holes and chaos}

\author[a,b,c]{Hyun-Sik Jeong,}
\author[d,e]{Keun-Young Kim,}
\author[d]{Gaya Yun}
\author[d]{and Hyeonwoo Yu}

\emailAdd{hyunsik.jeong@apctp.org}
\emailAdd{fortoe@gist.ac.kr}
\emailAdd{gayayun121@gm.gist.ac.kr}
\emailAdd{hyeonuyu759@gm.gist.ac.kr}

\preprint{{\large{\texttt{APCTP Pre2025 - 022}}}}

\affiliation[a]{Asia Pacific Center for Theoretical Physics, Pohang 37673, Korea}
\affiliation[b]{Department of Physics, Pohang University of Science and Technology, Pohang 37673, Korea}
\affiliation[c]{Instituto de F\'isica Te\'orica UAM/CSIC, Calle Nicol\'as Cabrera 13-15, 28049 Madrid, Spain}
\affiliation[d]{Department of Physics and Photon Science, Gwangju Institute of Science and Technology, \\ 123 Cheomdan-gwagiro, Gwangju 61005, Korea}
\affiliation[e]{Research Center for Photon Science Technology, Gwangju Institute of Science and Technology, \\ 123 Cheomdan-gwagiro, Gwangju 61005, Korea}

\abstract{
We study the quantum chaotic behavior of black holes within the brickwall model, focusing on probe scalar fields in ($d+1$)-dimensional hyperbolic AdS black holes. The brickwall model has captured the normal modes of BTZ black holes ($d=2$) with Gaussian-distributed boundary conditions on the stretched horizon and their connection to quantum chaos signatures of random matrix theory. Here, we extend this framework to higher-dimensional AdS black holes ($d>2$), exploring how black hole normal modes encode chaotic dynamics across dimensions and examining the universality of this approach. We show that quantum chaos is prominent in lower dimensions and persists in higher dimensions. Specifically, deviations from the logarithmic spectrum in $d=2$ evolves into a power-law spectrum at higher $d$, highlighting the sensitivity of black hole normal modes to dimensionality, while retaining signatures of quantum chaos despite the spectral deformation. Our results are supported by conventional diagnostics, including level spacing distributions and spectral form factors, as well as modern tools like Krylov complexity. Finally, we discuss the limitations of the brickwall model in capturing chaotic behavior in the parametrically large dimension limit, where the spectrum becomes constant leading to degeneracy and a non-chaotic regime, while emphasizing its effectiveness as a tool for studying quantum aspects of black holes in moderate higher dimensions.
}

\begin{document}
\maketitle

%
\section{Introduction}\label{}
Understanding the quantum nature of gravity remains one of the central challenges of modern high-energy physics, with black holes serving as pivotal testbeds for theoretical insights. The advent of the AdS/CFT correspondence, or more broadly the holographic duality~\cite{Maldacena:1997re,Gubser:1998bc,Witten:1998qj}, has profoundly deepened our knowledge of black hole physics. This framework has illuminated diverse phenomena, from microscopic entropy counting~\cite{Strominger:1996sh}, to the Hawking-Page transition~\cite{Witten:1998zw}, and to the longstanding puzzles of the information paradox and unitarity~\cite{Maldacena:2001kr}.

Within the broader context, 't Hooft's brickwall model~\cite{tHooft:1984kcu} offered an important step toward a microscopic understanding of black holes. By introducing a stretched horizon---effectively a Dirichlet wall placed at a Planckian distance outside the event horizon~\cite{Susskind:1993if}---and quantizing scalar field, the brickwall model provided a bulk description suggestive of a UV-complete description of a black hole microstates. More specifically, using statistical mechanics of the low-lying \textit{normal modes} of probe scalar fields, the model allowed for a derivation of black hole entropy through the partition function and density of states, thereby highlighting essential issues that quantum theory of gravity can address.\footnote{Recent analyses~\cite{Burman:2023kko,Krishnan:2023jqn} have shown that the normal modes can reproduce both Hawking temperature and entropy, once charges are specified, as a conventional statistical mechanics calculation.} While not a complete theory, the brickwall model has remained a guiding framework for further exploration of black hole thermodynamics and quantum mechanics.

From a more general perspective, it is widely believed that the spectrum of black hole microstates exhibits features characteristic of random matrix theory (RMT)~\cite{Cotler:2016fpe,Chen:2024oqv}. Recent progress has demonstrated this explicitly in the brickwall setting: for BTZ black holes, i.e., $(2+1)$ dimensional AdS black holes, imposing a Gaussian-distributed boundary condition on the stretched horizon yields \textit{normal mode statistics} that align with RMT signatures of quantum chaos~\cite{Das:2022evy,Das:2023ulz,Jeong:2024jjn}.\footnote{For pioneering discussions of brickwall models in AdS/CFT, see \cite{Kay:2011np,Iizuka:2013kma}, and for fermionic extensions in BTZ black holes, see \cite{Jeong:2024jjn}.} In these systems, scalar field excitations confined within the AdS black hole ``box" exhibit chaotic spectra, probed through conventional diagnostics such as level spacing distribution and spectral form factor, as well as more modern tools like Krylov complexity. 

Importantly, there are two key advancements beyond t' Hooft's original calculations in order to probe the quantum chaos signatures. First, the analysis of the brickwall model requires retaining the angular quantum numbers ($J$)~\cite{Das:2022evy,Das:2023ulz,Jeong:2024jjn}, which are typically discarded in black hole thermodynamic computations~\cite{tHooft:1984kcu,Burman:2023kko,Burman:2024egy}. In other words, the $J$-dependence in normal modes allows for an extension beyond thermal properties, enabling the capture of quantum chaotic dynamics. Additionally, the Gaussian-distributed boundary condition, inspired by angular-dependent profiles observed in BPS fuzzball microstates~\cite{Das:2023ulz}, constitutes a refinement of the original vanishing boundary condition on the stretched horizon~\cite{tHooft:1984kcu}.

It is worth noting that the conceptual motivation for the Dirichlet wall is from fuzzball constructions in string theory and supergravity~\cite{Mathur:2005zp}, where smooth horizons are replaced by stringy degrees of freedom localized near the would-be horizon. While explicit fuzzball solutions are technically intricate, the simpler brickwall model successfully captures some of their characteristic features.\footnote{See \cite{Krishnan:2023jqn,Burman:2023kko,Banerjee:2024dpl,Das:2024fwg,Burman:2024egy,Banerjee:2024ivh} for further recent analyses of normal modes. As the wall approaches the horizon, the normal modes condense rapidly, yielding an effective branch cut in the complex frequency plane.} In essence, the brickwall model provides a tractable framework to study quantum black holes and even their chaotic spectra within the universality class of RMT.\\

Despite all the advances on the quantum chaos of brickwall models with Gaussian distributions, most computations have been restricted to low-dimensional black holes, particularly BTZ black holes~\cite{Das:2022evy,Das:2023ulz,Jeong:2024jjn,Das:2023xjr,Ageev:2024gem}.\footnote{The normal modes with vanishing boundary conditions on the stretched horizon within four or five dimensional AdS black holes can be found in \cite{Krishnan:2023jqn,Das:2024fwg}.} For higher-dimensional AdS black holes, results still remain lacking, due to the difficulty of solving bulk equations of motion analytically for probe fields. In this manuscript, we aim to fill this gap by investigating the brickwall model for scalar fields in \textit{hyperbolic AdS black holes} across general dimensions. Specifically, we focus on AdS-Rindler geometries in $(d+1)$ dimensions with $d\geq2$, where the $d=2$ case reproduces the BTZ scenario.

The motivation or the guiding question of this manuscript is therefore, \textit{Does the brickwall model of AdS black holes, under Gaussian boundary conditions, universally exhibit quantum chaotic behavior in general dimensions?} We will show that in parametrically large dimensions, this may not be the case: the chaotic features can eventually disappear, which can also be understood from the quantization conditions of scalar fields in the limit of large dimension. Moreover, we demonstrate that the normal mode spectrum can deviate significantly from the logarithmic structure characteristic of BTZ black holes~\cite{Das:2023yfj,Basu:2025zkp} as the dimensionality increases.

Last but not least, it is also instructive to note that the hyperbolic space proves advantageous not only because it provides a manageable analytic framework for solving bulk equations, but also it provides supporting evidence for the AdS/CFT correspondence. Specifically, hyperbolic spaces serve as the setting for well-established holographic signatures of maximal chaos, such as the out-of-time-ordered correlator~\cite{Maldacena:2015waa,Sekino:2008he} and pole-skipping phenomena~\cite{Grozdanov:2017ajz,Blake:2017ris,Blake:2018leo}. These phenomena have been analyzed in AdS-Rindler geometries~\cite{Ahn:2019rnq,Ahn:2020bks,Ahn:2020baf} and successfully matched to CFT computations in hyperbolic space~\cite{Perlmutter:2016pkf}.\footnote{For recent developments on the out-of-time-ordered correlator and pole-skipping in holographic duality, see \cite{Ahn:2025exp}.} Therefore, investigating the brickwall model within hyperbolic space would provide valuable insights into the bulk dynamics with Dirichlet wall from the perspective of the boundary CFT, contributing to future studies in this area.\\

This manuscript is organized as follows.
In Section \ref{SEC2}, we provide a brief review of spectral diagnostics for quantum chaos: the level spacing distribution, spectral form factor, and the Krylov complexity of states. In Section \ref{SEC31} and \ref{SEC32}, we introduce the brickwall model of hyperbolic black holes in AdS spacetime and analyze the normal modes in generic dimensions. Section \ref{SEC33} presents the in-depth analysis of quantum chaos indicators derived from the normal modes discussed in Section \ref{SEC32}, and compares them with those obtained from random matrix theory. Finally, Section \ref{SEC4} is devoted to the conclusions.

%
\section{Preliminaries}\label{SEC2}
In this section, we provide a review of the spectral diagnostics of quantum chaos within the framework of random matrix theory, which will serve as a foundation for the analysis of normal modes in the subsequent sections. Our discussion includes the traditional indicators, such as level spacing distributions, as well as two closely related time-dependent measures: the spectral form factor and Krylov complexity of states. Readers familiar with these standard tools can skip directly to the following sections.

\subsection{Wigner surmise in random matrix theories}\label{}
Random matrix theory (RMT) is a powerful framework for describing the statistical behavior of complex systems, with broad applications in mathematics and physics~\cite{Guhr:1997ve,MEHTA1960395,Akemann:2011csh}. Central to RMT are the standard Gaussian ensembles---the symplectic (GSE), unitary (GUE), and orthogonal (GOE)---whose eigenvalue statistics have been extensively studied and are well established~\cite{Meh2004}.

A defining feature of RMT is its universality: spectral properties depend not on microscopic dynamics but on global symmetries that categorize systems into shared classes. In the context of quantum chaos, and in line with the Bohigas-Giannoni-Schmit conjecture~\cite{Bohigas:1983er}, RMT describes the statistical distribution of energy eigenvalues in quantum systems corresponding to classically chaotic counterparts.

Eigenvalues associated with different irreducible representations of the symmetry group are expected to be statistically independent and follow the appropriate ensemble statistics. Short-range fluctuations are most commonly analyzed using the nearest-neighbor spacing distribution $p(s)$, i.e., the probability density to find two adjacent (unfolded) energy levels with a normalized spacing $s$, while longer-range correlations are captured by measures such as the number variance or spectral rigidity.

When comparing spectra of different systems within the same symmetry class to predictions from RMT, it is necessary to apply an ``unfolding" procedure to the energy eigenvalues of both systems of interest and random matrices~\cite{Brody:1981cx}. This is to disentangle the non-universal part of the spectrum (e.g., the average behavior of the spectral density) from the universal  fluctuations that characterize spectral correlations. Practically, this involves a local rescaling of eigenvalues so that the unfolded spectrum has unit mean level spacing. The exact method of rescaling is not unique and can vary depending on the system under consideration.

\paragraph{Wigner surmise: the transition between GSE, GUE, and GOE.}
Nevertheless, for most practical purposes, it is convenient to employ the Wigner surmise~\cite{Neutron-Physics-by-Time-of-Flight:1956aa} rather than the exact analytical result of each RMT ensembles. It takes the form as
\begin{align}\label{WS}
\begin{split}
\text{Wigner surmise:} \quad p(s) = a(\beta) \, s^{\beta} \, e^{-b(\beta) s^2} \,, 
\end{split}
\end{align}
where the constants $a(\beta)$ and $b(\beta)$ are fixed by the conditions of normalization, 
\begin{align}\label{COND}
\begin{split}
\int_{0}^{\infty} p(s) \,\dd s = 1 \,, \qquad \langle{s}\rangle = \int_{0}^{\infty}  s \, p(s) \, \dd s = 1 \,,
\end{split}
\end{align}
which indicates that the total probability is unity and the mean level spacing equals one. The Wigner surmise \eqref{WS} yields the level spacing distributions of the GSE, GUE, and GOE~\cite{Wigner_1951,Dyson:1962oir,Dyson:1962es} for $\beta=4,2,1$, respectively. 

The key feature of the Wigner surmise is level repulsion, i.e., in chaotic systems, energy levels tend to avoid clustering. By contrast, integrable systems exhibit uncorrelated spectra described by a Poisson distribution,
\begin{align}\label{PS}
\begin{split}
\text{Poisson distribution:} \quad p(s) = e^{-s} \,, 
\end{split}
\end{align}
which allows arbitrary small spacing. This sharp contrast---correlated levels with repulsion in chaotic spectra versus uncorrelated levels in integrable ones---provides a fundamental diagnostic for distinguishing quantum chaos from integrability.

\paragraph{Brody distribution: an interpolation between GOE and Poisson.}
Physical systems often contain subsystems with distinct symmetries, or exhibit both integrable and chaotic dynamics. Varying a system parameter can therefore drive transition between different symmetry classes, in which case the Wigner surmise~\eqref{WS} provides a useful benchmark. Intermediate spectral statistics typically signal partial symmetry breaking or mixing or eigenvalues from subspaces.

A natural question is then whether such transitions can be captured as interpolations between RMT ensembles and Poisson statistics. Numerous studies confirm this picture. For instance, quantum billiards provide paradigmatic examples of the interplay between chaos and integrability, and several realizations exhibit GOE-Poisson transitions~\cite{Cheon:1991aa,Shigehara:1993aa,Csordas:1994aa,Abul-Magd:2008aa}. Similar behavior has been reported in the variants of SYK and stringy matrix models~\cite{Huh:2024ytz}, in the hydrogen atom under a magnetic field~\cite{Wintgen:1987aa}, and in condensed matter systems such as the  metal-insulator Anderson transition~\cite{Shklovskii:1993aa,Shukla_2005}, whose properties resemble Dyson's Brownian motion model~\cite{Dyson:1962brm}. These cases demonstrate the broad applicability of RMT to both pure and mixed systems.

To interpolate between the GOE and Poisson limits, Brody introduced a distribution~\cite{Brody1973} whose parameters are determined by least-squares fitting~\cite{Hsu:1997aa,Bos2007ParameterEF}:
\begin{align}\label{BD}
\begin{split}
\text{Brody distribution:} \quad p(s) =  (\beta+1) \, c(\beta) \, s^{\beta} \, e^{-c(\beta) \, s^{\beta+1}} \,,  \quad  c(\beta) = \Gamma\left(\frac{2+\beta}{1+\beta} \right)^{\beta+1} \,,
\end{split}
\end{align}
which satisfies the normalization conditions \eqref{COND}. This form interpolates smoothly between GOE $(\beta=1)$ and Poisson ($\beta=0$) statistics.\\

In essence, interpolations between RMT ensembles and Poisson can be described by the single parameter ($\beta$): using Wigner surmise \eqref{WS} for $\beta \in [4,1]$ (GSE-GOE) and the Brody distribution \eqref{BD} for $\beta \in [1,0]$ (GOE-Poisson). Although alternative interpolating forms exist, such as the Berry-Robnik distribution~\cite{Berry_1984}, empirical studies (e.g., in quantum billiards~\cite{Batisti__2013}) often find the Brody distribution provides superior fits despite its weaker theoretical foundation. In this manuscript, we adopt the Brody distribution to model the level statistics of the normal modes in the brickwall model in the following subsections.

\subsection{Time-dependent diagnostics of quantum chaos}\label{}
\paragraph{Spectral form factor and the ramp structure.}
Energy level statistics and spectral rigidity are the canonical static diagnostics of quantum chaos, directly reflecting the RMT hallmarks of chaotic spectra. Level repulsion reflects the suppression of nearly degenerate eigenvalues, while spectral rigidity encodes long-range correlations that enforce a universal RMT-like structure at late times. 

Complementing these static probes, the spectral form factor (SFF)~\cite{Brezin:1997aa,Cotler:2016fpe} provides a dynamical window into chaos. Its characteristic slope-dip-ramp-plateau structure illustrates the progression from early, system-specific dynamics to the universal late-time regime: the non-universal dip corresponds to short-time behavior, the \text{ramp} signals the onset of RMT correlations, and the plateau reflects eventual spectral saturation. The level spacing distribution, once unfolded, neatly captures this late-time universality.

The spectral form factor (SFF) is defined as 
\begin{align}\label{}
\begin{split}
\text{SFF} = \frac{|Z(\beta_T,t)|^2}{|Z(\beta_T,0)|^2}\,, \qquad Z(\beta_T,t) = \text{Tr} \left[e^{-(\beta_T-i t) H}\right]\,, 
\end{split}
\end{align}
where $\beta_T$ denotes the inverse temperature, $t$ the time, and $H$ the Hamiltonian of the quantum system of interest. In the context of brickwall model of black hole that will be studied in the next section, the normal mode frequencies $\omega$ of the black hole are identified as the eigenvalues of the quantum mechanical system. The partition function then takes the form:
\begin{align}\label{SFFFORBW}
\begin{split}
Z(\beta_T,t) = \sum_{\omega} e^{-(\beta_T-i t) \omega} \,.
\end{split}
\end{align}

\paragraph{Krylov complexity and the characteristic peak.}
Beyond traditional spectral probes, Krylov complexity $C(t)$ has been a modern diagnostics of quantum chaos~\cite{Balasubramanian:2022tpr,Caputa:2024vrn}. Originally formulated for operator growth in the Heisenberg picture~\cite{Parker:2018yvk}, it has been adapted to the Schr\"{o}dinger picture to measure the spread of states in Krylov subspace~\cite{Balasubramanian:2022tpr}. In this manuscript, we focus on the Krylov complexity of states. See \cite{Nandy:2024evd,Rabinovici:2025otw} for a detailed review, formalism, and a more complete list of references on Krylov complexity.

Krylov complexity is evaluated by constructing the Krylov basis $\{|K_n \rangle\}$ via the Lanczos algorithm~\cite{Lanczos:1950zz,RecursionBook} applied to the given Hamiltonian $H$. The recursion generates Lanczos coefficients $\{a_n,\,b_n\}$:
\begin{align}\label{}
    H|K_n \rangle = a_n | K_n \rangle + b_{n+1} | K_{n+1} \rangle + b_n | K_{n-1} \rangle \,,
\end{align}
with wave functions $\psi_n(t)$ evolving as 
\begin{align}\label{DES}
    i \, \partial_t \psi_n(t) = a_n \psi_n(t) + b_{n+1} \psi_{n+1}(t) + b_n \psi_{n-1}(t) \,.
\end{align}
The time-dependent state takes the form $|\psi(t) \rangle = \sum_n \psi_n(t) | K_n \rangle$, and Krylov complexity is then defined by
\begin{align}\label{eq:Krylov complexity}
    C(t) = \sum_{n} n \, |\psi_n(t)|^2 \,,
\end{align}
measuring the average spread of the state in Krylov space.

For thermofield double states taken as the initial state, an in particular for the maximally entangled states $\beta_T=0$, Krylov complexity $C(t)$ exhibits a four-stage profiles: initial growth, a characteristic peak, subsequent decay, and late-time saturation~\cite{Balasubramanian:2022tpr}. The appearance of the peak structure has been conjectured to be a robust hallmark of chaos, absent in integrable systems. Moreover, within this framework, $C(t)$ is fully determined by the underlying energy spectrum~\cite{Balasubramanian:2022tpr,Caputa:2024vrn}.

Notably, the peak of Krylov complexity effectively distinguishes chaotic from integrable phases, in agreement with conventional spectral diagnostics, level spacing statistics and spectral form factors, e.g., see \cite{Balasubramanian:2022tpr,Erdmenger:2023wjg,Hashimoto:2023swv,Camargo:2024deu,Baggioli:2024wbz,Balasubramanian:2024ghv,Huh:2024ytz,Baggioli:2025knt}, where it is also shown \cite{Huh:2024ytz} that the peak correlates with interpolating distributions such as Brody distribution Eq. \eqref{BD} in mixed phase space models like mass-deformed SYK models.

\paragraph{Interplay between SFF and Krylov complexity.}
The time evolution structure of Krylov complexity parallels the slope-dip-ramp-plateau feature of the SFF, establishing a conceptual bridge between late-time probes of chaos. At late times, their connection becomes quantitative~\cite{Erdmenger:2023wjg}: 
\begin{align}\label{SFFCRE}
\lim_{t\to\infty}\frac{1}{t}\int_{0}^{t}\mathrm{SFF}(\tilde{t}) \, \dd \tilde{t} = \frac{1}{1+2C(t \to \infty)} \,.
\end{align}
The saturation value of $C(t)$ directly fixes the late-time plateau of the SFF, where one can analytically find $C(t \to \infty) = N/2$ in the large system's size $N$ limit, since at late times, the Krylov amplitudes distribute uniformly as $|\psi_n(t)|^2 \approx 1/N$ with $\sum_{n=1}^{N}|\psi_n(t)|^2=1$. Note that $C(t \to \infty) = N/2$ simply implies that, at late times, the average position of a ``particle" in the Krylov subspace \eqref{eq:Krylov complexity} is at the midpoint of the Krylov dimension.

This direct relationship is reminiscent of how distinct diagnostics, spectral correlations (SFF) and dynamical spreading (Krylov complexity), converge to capture universal signatures of quantum chaos. For a comparative discussion of their characteristic timescales beyond the late-time regime, including Heisenberg time and system-size scaling, see \cite{Camargo:2024deu}.

%
\section{Brickwall model for hyperbolic AdS$_{d+1}$ black holes}\label{}

\subsection{Klein-Gordon equation in Rindler-AdS geometry}\label{SEC31}

\paragraph{Rindler-AdS spacetime.}
We begin by considering the $(d+1)$-dimensional Einstein-Hilbert action:
\begin{align}
S=  \int d^{d+1}x \sqrt{-g} \left[ R+\frac{d(d-1)}{\ell_{\text{AdS}}^2}\right] \,,
\end{align}
and, as a classical solution, we examine the Rindler-AdS$_{d+1}$ geometry, which is given by
\begin{align}\label{eq:RindlerAdS}
\dd s^2=-\left(\frac{r^2}{\ell_{\text{AdS}}^2}-1 \right) \dd t^2+\frac{\dd r^2}{\frac{r^2}{\ell_{\text{AdS}}^2}-1 }+r^2 \dd H_{d-1}^2 \,.
\end{align}
Here, $\ell_{\text{AdS}}$ denotes the AdS length scale, and $\dd H_{d-1}^2 = \dd \chi^2+\sinh^2\chi \dd \Omega_{d-2}^2$ is the line element of the $(d-1)$-dimensional hyperbolic space $H_{d-1}$, where $\dd \Omega_{d-2}$ is the line element of a unit $(d-2)$ sphere $S^{d-2}$. The black hole horizon is at $r=\ell_{\text{AdS}}$ and the AdS boundary is located at $r=\infty$. For simplicity, we set $\ell_{\text{AdS}}=1$ hereafter.

It is also convenient to introduce a new bulk radial coordinate $r = \cosh \nr$, in terms of which the metric~\eqref{eq:RindlerAdS} becomes
\begin{equation}\label{eq:RindlerAdS2}
\dd s^{2} = -\sinh^{2}\nr  \,\, \dd t^{2} + \dd \nr^{2} + \cosh^{2}\nr \,\, \dd H_{d-1}^{2} \,.
\end{equation}
Although the analysis can be generalized to any hyperbolic black hole with an arbitrary horizon radius different from $\ell_{\text{AdS}}$, we focus on the Rindler-AdS geometry \eqref{eq:RindlerAdS} or \eqref{eq:RindlerAdS2}, as in this case, the bulk equations of motion for probe fields can be solved analytically~\cite{Ahn:2019rnq,Ahn:2020bks,Ahn:2020baf}.\footnote{The hyperbolic black hole spacetime corresponds to a CFT in hyperbolic space $\mathbb{R} \times H_{d-1}$ and the maximally extended hyperbolic black hole geometry is dual to a TFD state. Notably, the AdS$_{d+1}$ geometry can be interpreted as an entangled state of a pair of CFTs on hyperbolic space~\cite{Czech:2012be,Perlmutter:2016pkf}, with an inverse temperature $2\pi \ell_{\text{AdS}}$, where the corresponding geometry is the Rindler-AdS geometry.}

\paragraph{Klein-Gordon equation in Rindler-AdS geometry.}
Next, we consider the action of the probe scalar field
\begin{align}
S_{\text{scalar}}=-\frac{1}{2}\int d^{d+1}x \sqrt{-g} \left(g^{\mu \nu} \partial_{\mu} \, \Phi \partial_{\nu} \Phi \right) \,,
\end{align}
where the field propagates on the background geometry \eqref{eq:RindlerAdS2}, and for simplicity, we focus on the massless case. Then, the corresponding equation of motion is given 
\begin{equation}
\label{eq:klein}
\frac{1}{\sqrt{-g}}\,\partial_\mu\!\left(\sqrt{-g}\, g^{\mu\nu}\,\partial_\nu \Phi\right) = 0 \,.
\end{equation}
In terms of the coordinates $(t,\nr,x^i)$, where $x^i \in \mathbb{H}^{d-1}$, it can be further expressed as
\begin{align} \label{eq-scalar}
\partial_{\nr}^2 \Phi -\frac{\partial_t^2 \Phi}{\sinh^2 \nr}+\frac{\square_{H^{d-1}} \Phi }{\cosh^2 \nr}+ \big[ \coth \nr+(d-1) \tanh \nr \big] \partial_{\nr} \Phi =0\,,
\end{align}
where 
\begin{align} \label{}
\square_{H^{d-1}}= \partial_{\chi}^2+(d-2) \coth \chi \partial_{\chi}+\frac{1}{\sinh^2\chi} \square_{S^{d-2}} \,,
\end{align}
is the Laplacian operator in $\mathbb{H}^{d-1}$.

To solve this equation, we use the Fourier transformation
\begin{align}
\Phi(t,\nr,\chi)= \int \dd \omega \dd J \, \phi(\nr; \omega, J) \, e^{-i \omega t} \, \mathbb{S}_{J}(\chi)\label{ssh}\,,
\end{align}
where $\mathbb{S}_{J}(\chi)$ is an eigenfunction of $\square_{H_{d-1}}$, defined by the equation
\begin{align}\label{YLM}
\left[\square_{H_{d-1}} + J^2 + \frac{(d-2)^2}{4} \right] \mathbb{S}_{k}(\chi) = 0 \,.
\end{align}
For more detailed derivation using hyperspherical harmonics, see \cite{Ahn:2019rnq,Ahn:2020bks,Ahn:2020baf}. In terms of this Fourier mode, $\phi(\nr; \omega, J)$, we have the equation of motion \eqref{eq-scalar} as
\begin{align} \label{eq-F}
\phi''(\nr)+\big[ \coth \nr+(d-1) \tanh \nr \big] \phi'(\nr)+\left[\frac{\omega^2}{\sinh^2\nr}-\frac{J^2 + \frac{(d-2)^2}{4}}{\cosh^2\nr}  \right] \phi(\nr)=0\,,
\end{align}
where we omit the $\omega$ and $J$ dependence in $\phi(\nr; \omega, J)$ for clarity and brevity. In the context of brickwall model, we now assume that $J$ represents the angular quantum number~\cite{tHooft:1984kcu,Das:2022evy,Das:2023ulz,Jeong:2024jjn}.

\subsection{Normal modes of probe scalar fields}\label{SEC32}
\paragraph{Brickwall models in generic dimensions.}
We now proceed with computing the normal modes of hyperbolic black holes within the brickwall framework. Specifically, we extend the analysis conducted in the $d=2$ case in references~\cite{tHooft:1984kcu,Das:2022evy,Das:2023ulz,Jeong:2024jjn} to involve arbitrary dimensions. 

We start with the bulk equation \eqref{eq-F}, which has an analytic solution given by:
\begin{align}\label{eq:Fz}
\begin{split}
\phi(z) = e^{-\tfrac{\pi\omega}{2}}\, z^{-\tfrac{i\omega}{2}}
\Bigg[&
C_{1}\, e^{\pi\omega}\, 
{}_2F_{1}\!\left(
\tfrac{2-d+2 i (J-\omega)}{4} ;
\tfrac{2-d-2 i (J+\omega)}{4}  ;
1-i\omega ; z
\right) \\
& + C_{2}\,  z^{i\omega}\, 
{}_2F_{1}\!\left(
\tfrac{2-d-2 i (J-\omega)}{4} ;\,
\tfrac{2-d+2 i(J+\omega)}{4} ;\,
1+i\omega;\, z
\right)
\Bigg]
\end{split}
\end{align}
where the coordinate transformation~\cite{Jeong:2024jjn} simplifies the subsequent computation of normal modes
\begin{align}\label{zcoor}
\begin{split}
z=\tanh^{2} \nr \,,
\end{split}
\end{align}
with $z=0$ corresponding to the horizon and $z=1$ representing the AdS boundary.

Next, to investigate the normal modes, first we analyze the solution near the AdS boundary ($z=1$):
\begin{align}\label{}
\begin{split}
\phi_{\text{bdry}}(1) \approx  C_{1}\,\mathcal{A}(\omega,J,d) + C_{2}\,\mathcal{B}(\omega,J,d) \,,
\end{split}
\end{align}
where
\begin{equation}\label{eq:AB-def}
\begin{aligned}
\mathcal{A}(\omega,J,d)
&=\frac{e^{\pi\omega}\,\Gamma(1-i\omega)}
{\Gamma\!\big[\tfrac14(2+d+2i(J-\omega))\big]\,
 \Gamma\!\big[\tfrac14(2+d-2i(J+\omega))\big]} \,, \\
\mathcal{B}(\omega,J,d)
&=\frac{\Gamma(1+i\omega)}
{\Gamma\!\big[\tfrac14(2+d-2i(J-\omega))\big]\,
 \Gamma\!\big[\tfrac14(2+d+2i(J+\omega))\big]} \,.
\end{aligned}
\end{equation}
Enforcing the normalizability condition~\cite{Das:2022evy,Das:2023ulz,Jeong:2024jjn}, $\phi_{\text{bdry}}(1)=0$, we find the relationship between $C_2$ and $C_1$:
\begin{align}\label{C2C1}
\begin{split}
C_{2} = -\,C_{1}\,\frac{\mathcal{B}(\omega,J,d)}{\mathcal{A}(\omega,J,d)} \,.
\end{split}
\end{align}
This condition is the standard sourceless condition for quasi-normal modes in AdS black holes~\cite{Horowitz:1999jd,Birmingham:2001pj,Son:2002sd,Kovtun:2005ev,Berti:2009kk,Berti:2025hly}.

In the brickwall model, we also impose a boundary condition at the stretched horizon ($z=z_0$), just outside the event horizon ($z=0$). Plugging $C_2$ from the previous result \eqref{C2C1} into the bulk equation \eqref{eq:Fz}, we find the radial solution near the horizon, a linear combination of both incoming and outgoing solutions:
\begin{equation}
\phi_{\text{hor}}(z) \approx C_{1}\left( P_{1}\, z^{-\tfrac{i\omega}{2}} + Q_{1}\, z^{\tfrac{i\omega}{2}} \right) \,,
\end{equation}
where 
\begin{equation}
\label{eq:PQ-def}
P_{1} \;=\; 1, 
\quad
Q_{1} \;=\; -\,\frac{
\Gamma\!\left[1 - i\omega\right]\,
\Gamma\!\left[\tfrac{1}{4}\left(2+d-2 i (J-\omega)\right)\right]\,
\Gamma\!\left[\tfrac{1}{4}\left(2+d+2 i(J+\omega)\right)\right]
}{
\Gamma\!\left[1 + i\omega\right]\,
\Gamma\!\left[\tfrac{1}{4}\left(2+d+2 i (J-\omega)\right)\right]\,
\Gamma\!\left[\tfrac{1}{4}\left(2+d-2 i(J+\omega)\right)\right]
} \,,
\end{equation}
with  $|P_{1}| = |Q_{1}|$.  Note that the form of $P_1$ are simplified to unity due to the coordinate transformation \eqref{zcoor}. \eqref{eq:PQ-def} is the extension of the $d=2$ case given in \cite{Jeong:2024jjn}.

The boundary condition at the stretched horizon as the Dirichlet condition~\cite{Das:2023ulz} is then expressed as
\begin{align}\label{ST0}
\begin{split}
\phi_{\text{hor}}(z=z_0) = C_1 \left( P_1 \, z_0^{-\frac{i\omega}{2}} + Q_1 \,  z_0^{\frac{i\omega}{2}}  \right) =: \mu \, e^{i \lambda \,\omega} \,.
\end{split}
\end{align}
This condition is motivated by the angle-dependent profiles found in BPS fuzzballs. Next, we proceed by introducing phase terms for $P_1$ and $Q_1$
\begin{align}\label{}
\begin{split}
P_1 = |P_1| e^{i \theta_{P}}  \,, \qquad  Q_1 = |Q_1| e^{i \theta_{Q}} \,.
\end{split}
\end{align}
Substituting these into the boundary condition \eqref{ST0}, we obtain
\begin{align}\label{ST2}
\begin{split}
e^{i (\theta_{P}-\theta_{Q})} = \mu \, e^{i \left(\lambda \, \omega + \frac{\theta}{2}\right)} - e^{i \theta}  \,, 
\end{split}
\end{align}
where $\theta = \text{Arg} \left[z_0^{i\omega}\right]$. The remaining freedom is fixed by setting $C_1 Q_1 =1$. From this, the real and imaginary parts yield $\mu = 2 \cos \left(\lambda \, \omega - {\theta}/{2}\right)$, and the quantization conditions on $\omega$ are given by
\begin{align}\label{ST3}
\begin{split}
\text{Quantization condition:} \quad\, \cos (\theta_{P}-\theta_{Q}) = \cos (2 \lambda\,\omega) \,, \quad \sin (\theta_{P}-\theta_{Q}) = \sin (2 \lambda\,\omega) \,.
\end{split}
\end{align}
These phase equations allow us to compute the normal modes $\omega(n\,,J)$, where $n$ is an integer quantum number.

The normal mode $\omega(n\,,J)$ can be determined by solving the quantization conditions \eqref{ST3}, with (I) the phases $\theta_{P}$ and $\theta_{Q}$, derived from the analytic bulk solutions \eqref{eq:Fz}, and (II) the parameter $\lambda$.

Regarding $\lambda$, as discussed in \cite{Das:2022evy,Das:2023ulz,Jeong:2024jjn}, it can be modeled as drawn from a \textit{Gaussian distribution}. The Gaussian ensemble approach provides a controlled framework for modeling randomness, facilitating comparisons with predictions from random matrix theory. This method is consistent with the effective description of chaotic systems and random matrix quantum mechanics.

In addition, we set $\langle \lambda \rangle = \frac{1}{2} \log z_0$ by ensuring that $\mu=2$ in the zero-variance limit, with $\lambda$ varying according to the Gaussian distributions's variance $\sigma$. Notably, $\langle \lambda \rangle$ is heuristically comparable to the position of the stretched horizon, and as $z_0 \rightarrow 0$, $\langle \lambda \rangle\rightarrow-\infty$.

\paragraph{Normal modes of scalar fields in higher dimensions.}
In this manuscript, we set $\langle \lambda \rangle=-10^4$, as in the BTZ cases~\cite{Das:2022evy,Das:2023ulz,Jeong:2024jjn}, to compute the normal modes. Four remarks are in order: (I) the redefinitions of $\lambda$ in \eqref{ST0}, such as $\lambda \rightarrow\lambda/\omega$, does not affect the physical properties on the normal modes; (II) we focus on the first non-zero mode $\omega(n=1,J)$, denoted $\omega(J)$ to avoid clutter, as higher modes are less significant in the statistical analysis of normal mode spectra; (III) the features of quantum chaos with the brickwall model become more pronounced as the stretched horizon approaches the event horizon, where $\langle \lambda \rangle=-10^4$ is sufficient to exhibit chaotic features; (IV) while various choices for the variance $\sigma$ are conceivable such as $\sigma=\sigma_0,\, \sigma_0/J,\, \sigma_0/\sqrt{J}$, the statistical properties remains unchanged. For the sake of simplicity, we adopt $\sigma=\sigma_0$ without any scaling, and henceforth omit the subscript for clarity.

\begin{figure}[]
\centering
     \subfigure[$\sigma=0.0022$ (GUE)]
     {\includegraphics[width=7.0cm]{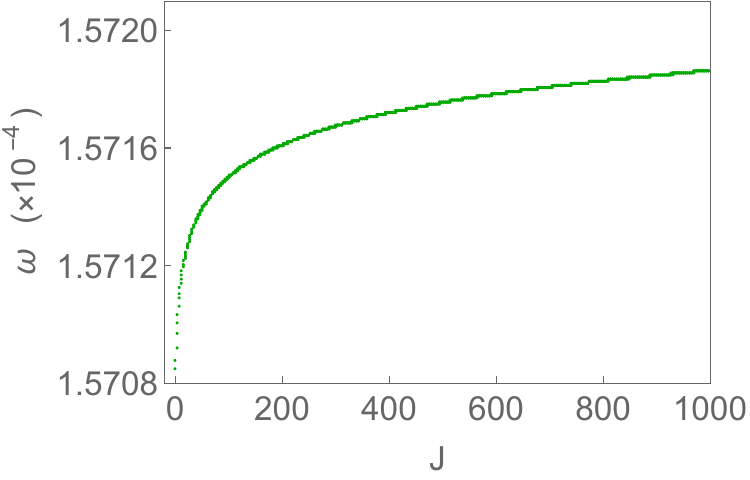} \label{}}
\quad
     \subfigure[$\sigma=0.2$ (Poisson)]
     {\includegraphics[width=7.0cm]{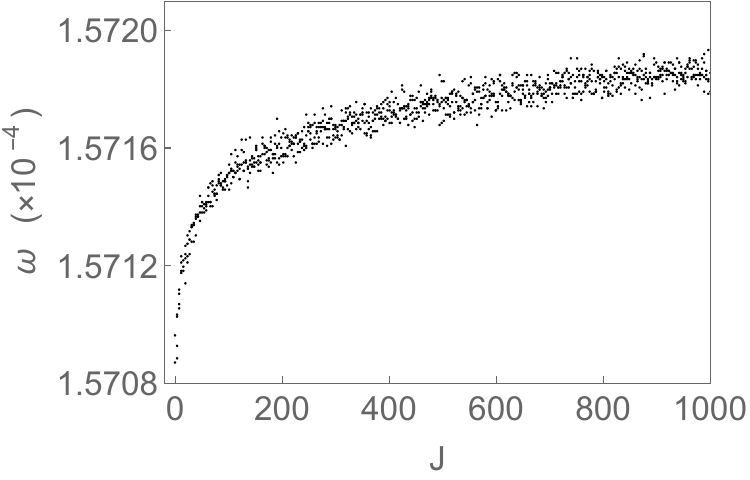} \label{}}
 \caption{Normal mode spectrum for $d=3$ (AdS$_4$) with $\sigma = 0.0022$ (left) and $0.2$ (right).}\label{DRAFTFIG1}
\end{figure}

We calculate the normal modes for scalar fields with various values of the variance $\sigma$ at a fixed dimension $d$, as outlined in the procedure above. Representative higher-dimensional results for the normal mode spectrum when $d=3$ (AdS$_4$) are shown in Fig. \ref{DRAFTFIG1}, where we present the normal mode spectrum for two different values of $\sigma$.

As the variance increases, the normal mode spectrum $\omega(J)$ becomes highly erratic. In the following section, we demonstrate that for $\sigma = 0.0022$, the level spacing distribution follows the statistics of the GUE ensemble, where for $\sigma = 0.2$, it transitions to a Poisson distribution. In addition, we will show that the distribution can be fitted with GSE, GUE, GOE, and Poisson statistics through \eqref{WS} and \eqref{BD}, depending on more specified values of $\sigma$ in higher dimensions. These findings imply that the brickwall model framework can be effective in investigating chaos even in higher-dimensional spacetimes, extending beyond the BTZ black holes in $d=2$ dimensions~\cite{Jeong:2024jjn}.

\begin{figure}[]
\centering
     {\includegraphics[width=7.2cm]{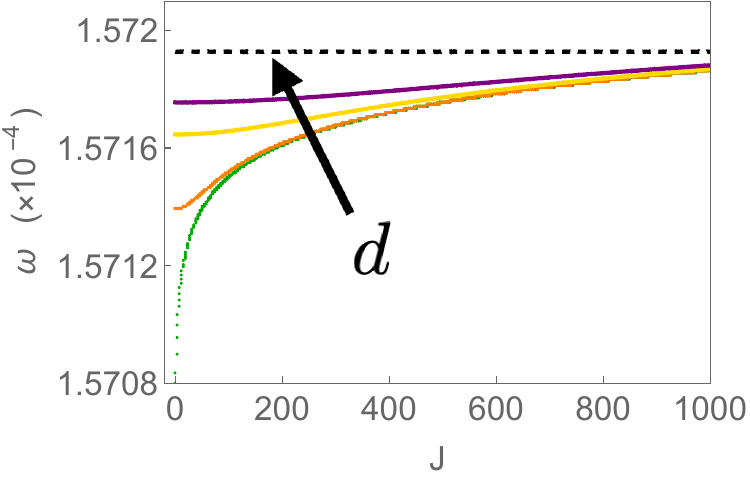} \label{}}
 \caption{Normal mode spectrum for $\sigma = 0.0022$ with $d=2, 100, 500, 1000$ (green, orange, yellow, purple). The black dashed line is the analytic result from \eqref{ANDLIMIT} for $d=3000$, giving $\omega = 0.000157193$.}\label{DRAFTFIG2}
\end{figure}

We further discuss the spatial dimension effect on the normal mode spectrum with a fixed $\sigma$. See Fig. \ref{DRAFTFIG2}.
We find that for $d=2$, the normal mode spectrum exhibits a logarithmic form at small $J$ regime, with $\omega \approx \log J$, while for higher dimensions, the spectrum transitions into a power-law form (e.g., for $d=500$, the data fits $\omega \approx J^2$). This power-law behavior is a novel observation and marks a deviation from the previously expected logarithmic spectrum in lower dimensions~\cite{Das:2022evy,Das:2023ulz,Jeong:2024jjn}. It explicitly highlights the sensitivity of black hole normal modes to dimensionality.

\paragraph{More on the normal modes in the large $d$ limit.}
It is instructive to note that in the extremely large $d$ limit ($d\rightarrow\infty$), the normal mode frequency approaches a constant, which can be understood from $Q_1$ in \eqref{eq:PQ-def}. In this limit, $Q_1$ becomes
\begin{align}\label{ANDLIMIT}
Q_1 \approx -2^{-4 i \omega}\, d^{2 i \omega} \, \frac{\Gamma(1-i\omega)}{\Gamma(1+i\omega)}  \,.
\end{align}
This indicates that in the parametrically large $d$ limit, the normal mode frequency $\omega$ becomes independent of $J$, which is in agreement with the numerical findings presented in Fig. \ref{DRAFTFIG2}. Consequently, the spectrum degenerates, with multiple states sharing the same mode. Such degeneracy suggests a deviation from chaotic dynamics, as will be demonstrated shortly.

The analysis of the limit where the spatial dimension is taken to infinity is a familiar strategy in statistical mechanics, e.g., infinite-coordination lattices render mean-field theory exact~\cite{Georges:1996zz}, but less common in continuum field theories. In quantum field theory, the limit is generally ill-defined due to ultraviolet divergences, yet classical field theories, notably General Relativity, can remain well behaved. 

In fact, gravitational systems simplify at large $d$, permitting analytic control of black hole quasinormal modes and more tractable simulations of black hole dynamics~\cite{Emparan:2013moa}. These developments have direct implications for simplified numerical simulations of black and gravitational wave physics. See~\cite{Emparan:2020inr} for a comprehensive review including the discussion of generic aspects of black holes in this large $d$ limit with the applied AdS/CFT perspective.

A key result in this program is the discovery of two distinct dynamical regimes in large-$d$ black holes, with quasinormal frequencies separated parametrically in $1/d$. Many such modes fall into broad universality classes determined only by horizon radius or shape~\cite{Emparan:2013moa,Emparan:2013xia} (i.e., classes with no information about the black hole~\cite{Emparan:2020inr}). In this spirit, our finding that the normal mode spectrum saturates to a constant value may resemble the disappearance of some features of black hole: microstate information is washed out, and the spectrum reduces to a simple degenerate structure. The usual random-matrix-like features of black hole microstates, \textit{the} common lore, may disappear in this regime.

Thus, whilst the large-$d$ limit provides an analytically controlled framework in which degeneracy replaces chaos, and the dependence on systems's parameters such as $J$ vanishes. A deeper physical interpretation of this phenomena remains an open and intriguing direction beyond the scope of this work.

\subsection{Spectral statistics of normal modes and quantum chaos}\label{SEC33}
Next, we provide the statistical spectral analysis building on the normal modes of the scalar field derived above, where modes are interpreted as eigenvalues of scalar excitations. 

Especially, we examine their statistical properties and compare them with those of random matrices, using the chaos diagnostics outlined in Section \ref{SEC2}: level spacing distribution, spectral form factor (SFF), and Krylov complexity. Our study is carried out in the infinite temperature limit, as in \cite{Das:2022evy,Das:2023ulz,Jeong:2024jjn}, where the thermofield double state is maximally entangled. This limit provides a clear framework for connecting SFF and Krylov complexity \eqref{SFFCRE}.

As shown earlier, increasing the spatial dimension $d$ drives the spectrum from a logarithmic to a power-law scaling: see Fig. \ref{DRAFTFIG2}. To further prove the robustness of brickwall models in capturing signatures of quantum chaos in higher dimensions, we mainly focus on $d=2,100,500$: the $d=2$ case exemplifies the logarithmic regime, while the higher-dimensional cases exhibit power-law spectra.

\subsubsection{Level spacing distribution}\label{}
\begin{figure}[]
\centering
     \subfigure[$\sigma=0.0015$ (GSE)]
     {\includegraphics[width=4.4cm]{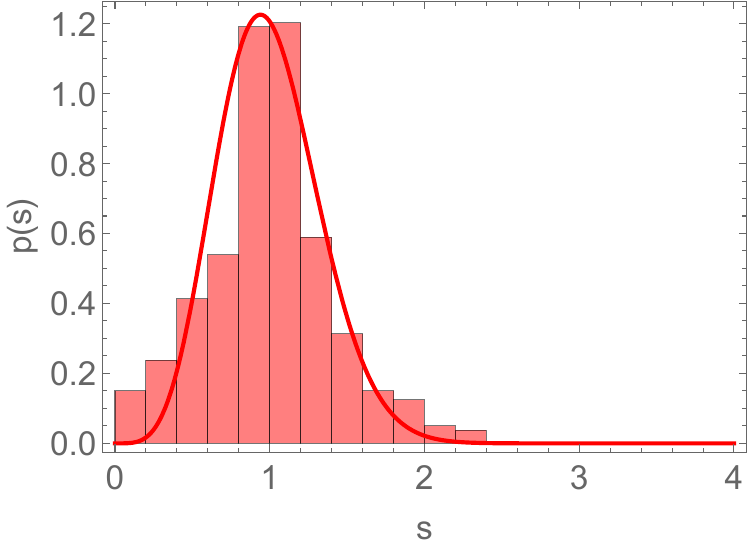} \label{}}
\quad
     \subfigure[$\sigma=0.0022$ (GUE)]
     {\includegraphics[width=4.4cm]{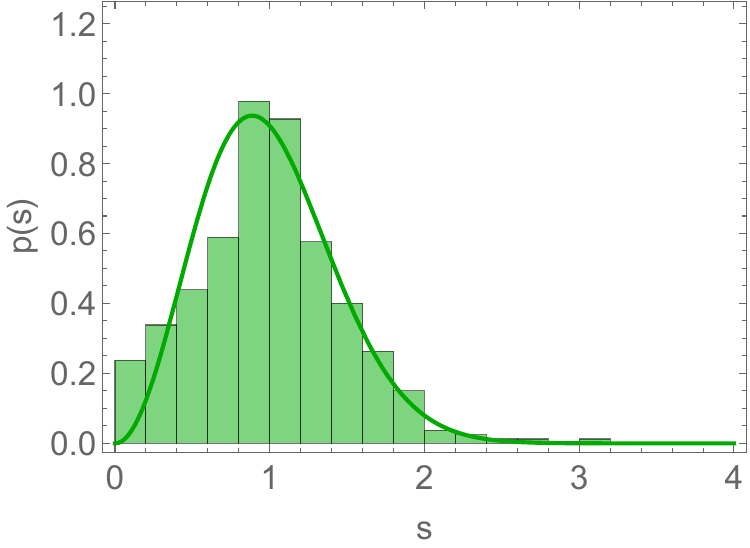} \label{}}
\quad
     \subfigure[$\sigma=0.0028$ (GOE)]
     {\includegraphics[width=4.4cm]{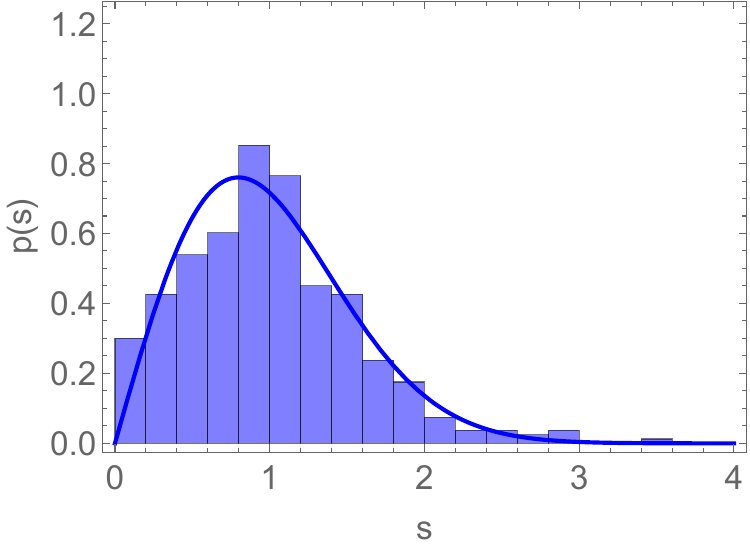} \label{}}
     
     \subfigure[$\sigma=0.0011$ (GSE)]
     {\includegraphics[width=4.4cm]{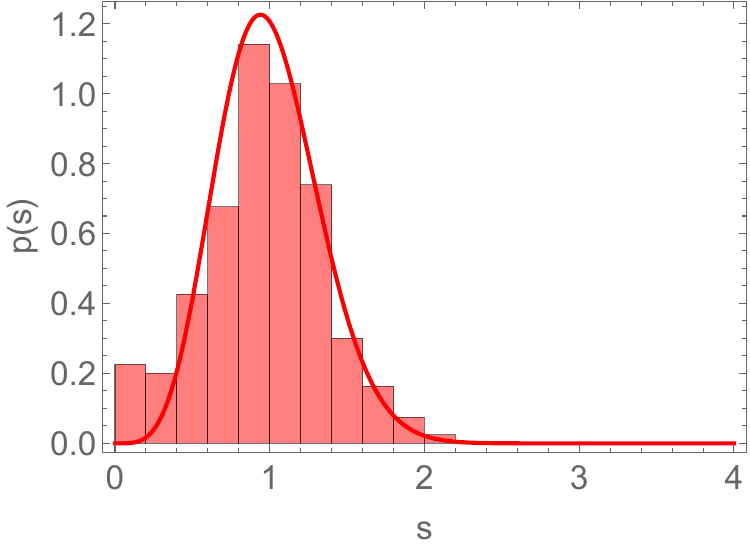} \label{}}
\quad
     \subfigure[$\sigma=0.0016$ (GUE)]
     {\includegraphics[width=4.4cm]{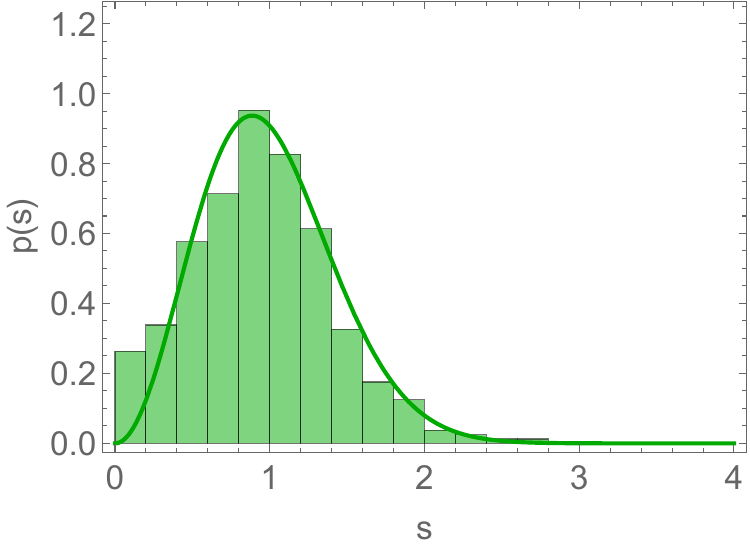} \label{}}
\quad
     \subfigure[$\sigma=0.0020$ (GOE)]
     {\includegraphics[width=4.4cm]{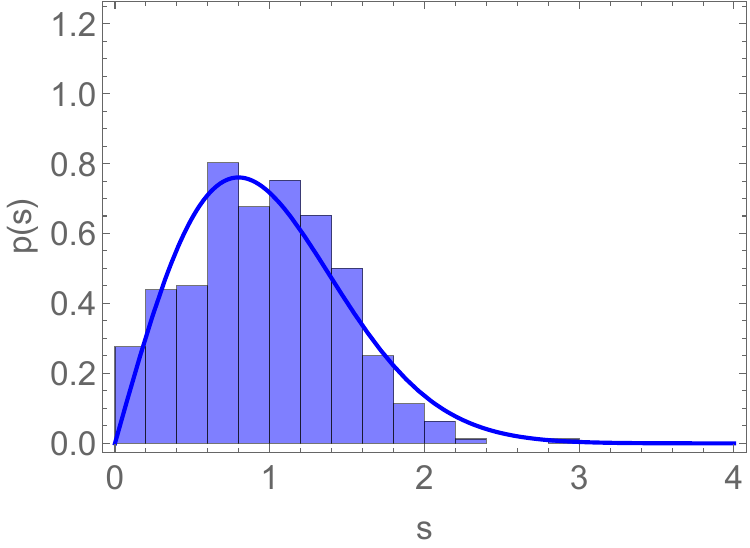} \label{}}
     
     \subfigure[$\sigma=0.0004$ (GSE)]
     {\includegraphics[width=4.4cm]{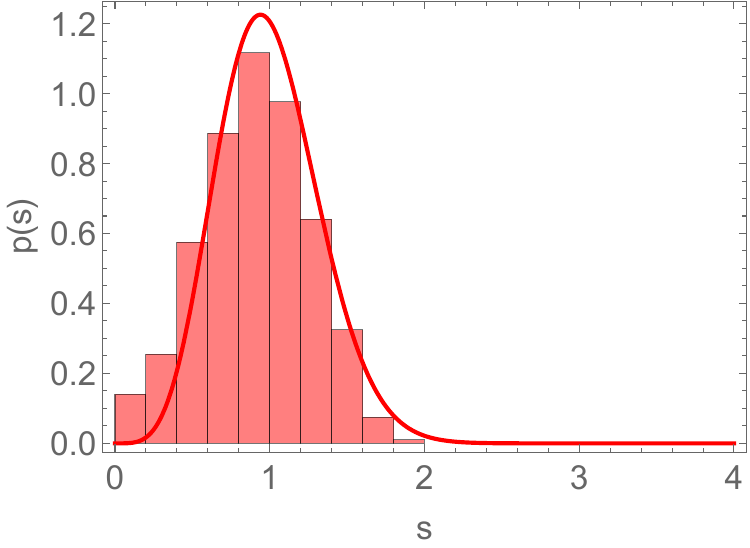} \label{}}
\quad
     \subfigure[$\sigma=0.0005$ (GUE)]
     {\includegraphics[width=4.4cm]{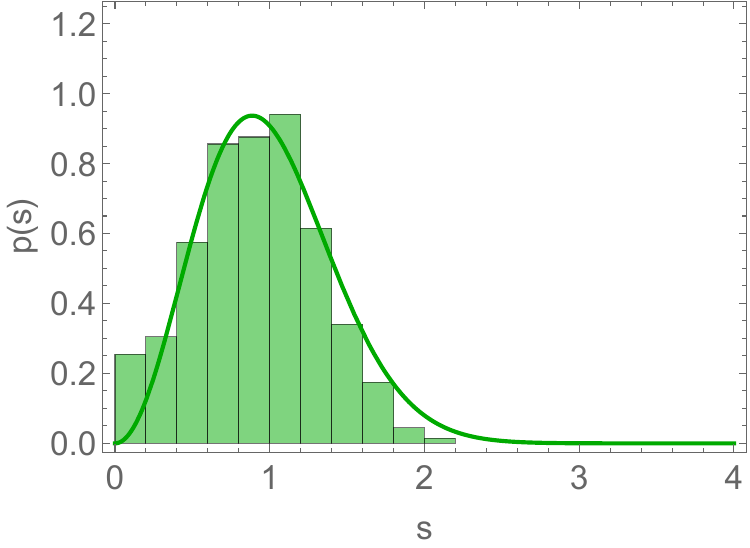} \label{}}
\quad
     \subfigure[$\sigma=0.0007$ (GOE)]
     {\includegraphics[width=4.4cm]{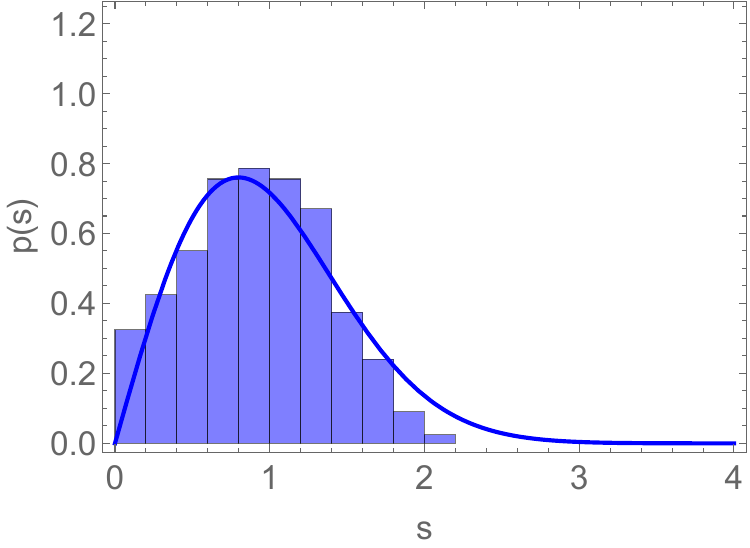} \label{}}
 \caption{Level spacing distributions of scalar fields for $d=2$ (a-c), $d=100$ (d-f), and $d=500$ (g-i). The solid curves represent the Wigner surmise from random matrix theory \eqref{WS}, corresponding to the GSE ($\beta=4$), GUE ($\beta=2$), and GOE ($\beta=1$).}\label{DRAFTFIG3}
\end{figure}

Figure \ref{DRAFTFIG3} shows the level spacing distributions of scalar fields across different dimensions. The brickwall model reproduces all RMT classes in both low and high dimensions, matching the Wigner surmise \eqref{WS} for GSE ($\beta=4$), GUE ($\beta=2$), and GOE ($\beta=1$), when the variance is set to $\sigma=0.0015, 0.0022, 0.0028$ respectively for $d=2$. The values of $\sigma$ depend on the dimension, as further illustrated in Fig. \ref{DRAFTFIG6}. 

These results confirm that signatures of quantum chaos within the brickwall models appear not only in the logarithmic spectrum at $d=2$ but also in the power-law spectra characteristic of higher dimensions.
\begin{figure}[]
\centering
     \subfigure[$\sigma=0$]
     {\includegraphics[width=6.2cm]{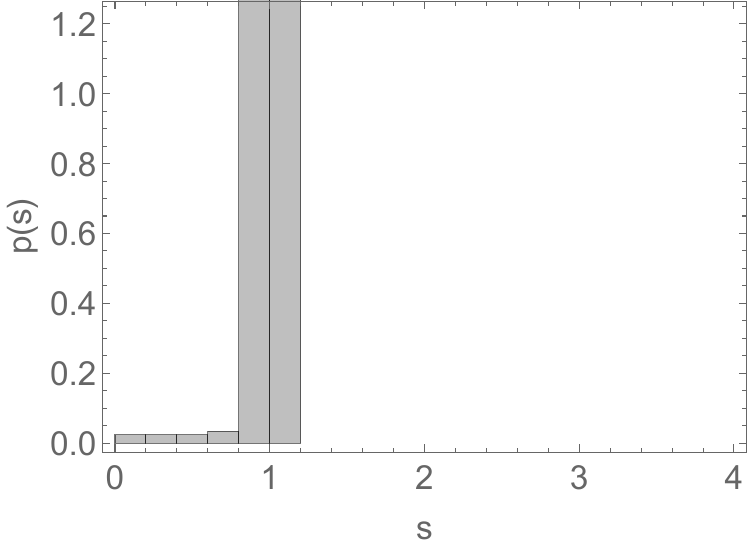} \label{}}
\quad
     \subfigure[$\sigma=0.2$]
     {\includegraphics[width=6.2cm]{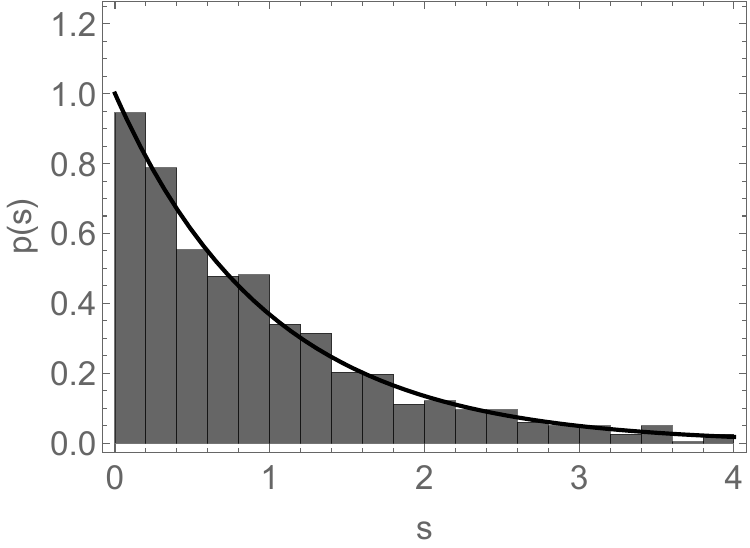} \label{}}
 \caption{Level spacing distributions of scalar fields for $d=100$ with $\sigma=0$ (left) and $0.2$ (right), where the black solid line is the Poisson distribution \eqref{PS}.}\label{DRAFTFIG4}
\end{figure}

Before analyzing generic cases spanning the broader values of $\sigma$, we revisit the extremal limits discussed in \cite{Das:2022evy,Das:2023ulz,Jeong:2024jjn}. For $\sigma=0$, the spectrum collapses to a sharp peak approximating a delta function, while at $\sigma=0.2$ the distribution approaches the Poisson form \eqref{PS}, as shown in Fig. \ref{DRAFTFIG4} for $d=100$. These features persist across dimensions.

For $d=2$, \cite{Jeong:2024jjn} argued that the $\sigma=0$ case resembles saddle-dominated scrambling systems: although integrable, it exhibits signatures of chaotic-like dynamics. This interpretation is supported by the delta-like level spacing distribution, the steep ramp in the spectral form factor, and the peak structure in Krylov complexity. We demonstrate that this behavior extends to higher dimensions.

The pronounced dependence of spectral statistics on small variations in $\sigma$ appears to be an intrinsic property of the model rather than a technical artifact. This behavior mirrors transitions observed in quantum chaotic systems, highlighting the brickwall model as an effective framework for probing universality and the breakdown of integrability in curved spacetime.

\begin{figure}[]
\centering
     {\includegraphics[width=6.8cm]{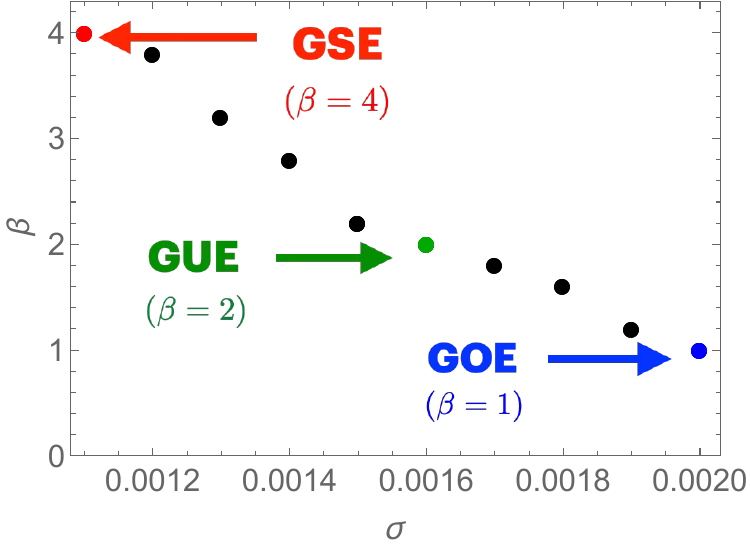} \label{}}
\quad
     {\includegraphics[width=7.0cm]{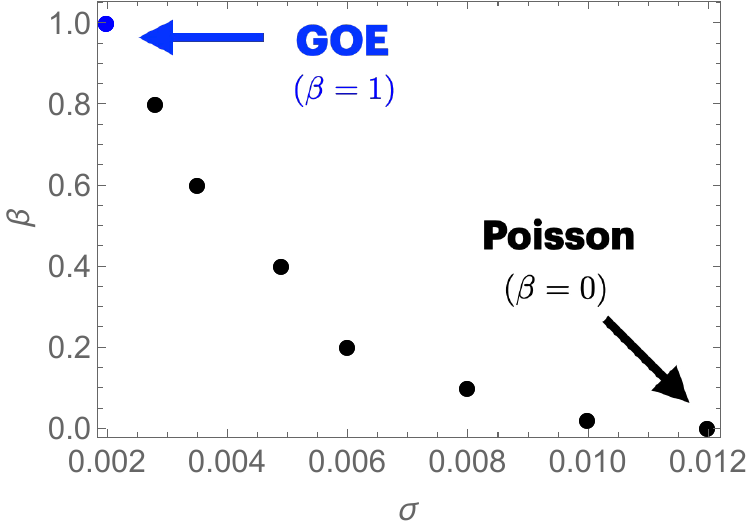} \label{}}
 \caption{Dependence of the level spacing distribution parameter $\beta$ on $\sigma$ for $d=100$: Wigner surmise \eqref{WS} (left) and Brody distribution \eqref{BD} (right).}\label{DRAFTFIG5}
\end{figure}

\begin{figure}[]
\centering
     {\includegraphics[width=7.2cm]{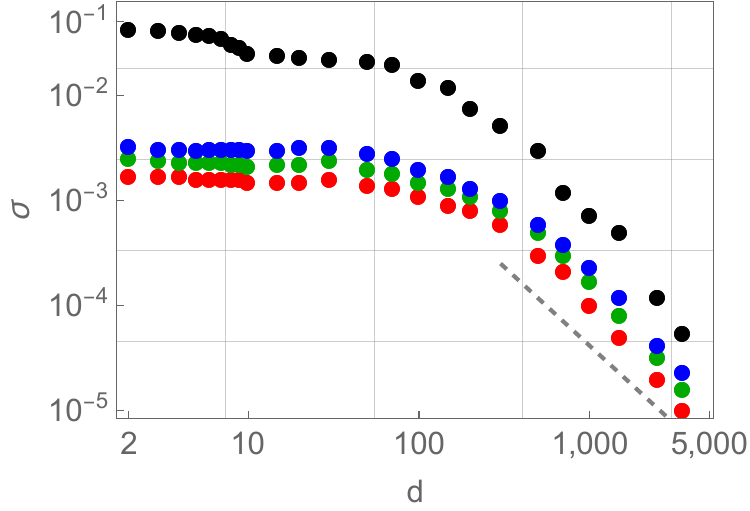} \label{}}
 \caption{Variance $\sigma$ as a function of dimension $d$ for GSE, GUE, GOE, and Poisson (red, green, blue, black). The dashed line represents the fit $\sigma\approx1/d^{1.5}$.}\label{DRAFTFIG6}
\end{figure}

To substantiate this interpretation, we extend the analysis beyond the fundamental RMT ensembles (GSE, GUE, GOE) by exploring a wider range of $\sigma$ and comparing the resulting level statistics with both the Wigner surmise \eqref{WS} and the Brody distribution \eqref{BD}. As shown in Fig. \ref{DRAFTFIG5}, the single distribution parameter $\beta$, interpreted as the Dyson index or Brody parameter, varies monotonically with $\sigma$.

Finally, we conclude the level-statistics analysis by examining the dimension dependence of $\sigma$ in Fig. \ref{DRAFTFIG6}, which collects the $\sigma$ values corresponding to RMT ensembles and the Poisson distribution for each dimension. The results indicate that the fundamental RMT classes persist in higher dimensions. However, in the limit $d \rightarrow \infty$, the characteristic signatures of chaos progressively disappear, as the scaling $\sigma\approx1/d^{1.5}$ implies that no RMT ensemble can be reliably identified for finite values of $\sigma$. This behavior is consistent with the description around Eq. \eqref{ANDLIMIT}, where the degeneracy of normal modes becomes manifest.\footnote{In the limit $d \rightarrow \infty$, the level spacing distribution is expected to converge toward a combination of two limiting forms: the delta-function-like distribution at $\sigma=0$, and a Poisson distribution characterized by $p(0)=1$, corresponding to uncorrelated spectra. Thus, in the regime of infinite dimensionality, one may heuristically expect the distribution to exhibit a delta peak at $s=0$, indicative of purely degenerate energy levels.}

\subsubsection{Spectral form factor and Krylov complexity of states}\label{}
We now focus on the values of $\sigma$ corresponding to GSE, GUE, GOE, and Poisson level statistics given in Fig ~\ref{DRAFTFIG3}, and employ them consistently in time-dependent chaos diagnostics. In essence, we show that both the spectral form factor (SFF) and the Krylov complexity yield results in agreement with the level-spacing analysis, reinforcing the robustness of the brickwall model as a tool for probing quantum chaotic features of black holes in higher dimensions.

\begin{figure}[]
\centering
     \subfigure[$\sigma=0.0015$ (GSE)]
     {\includegraphics[width=4.4cm]{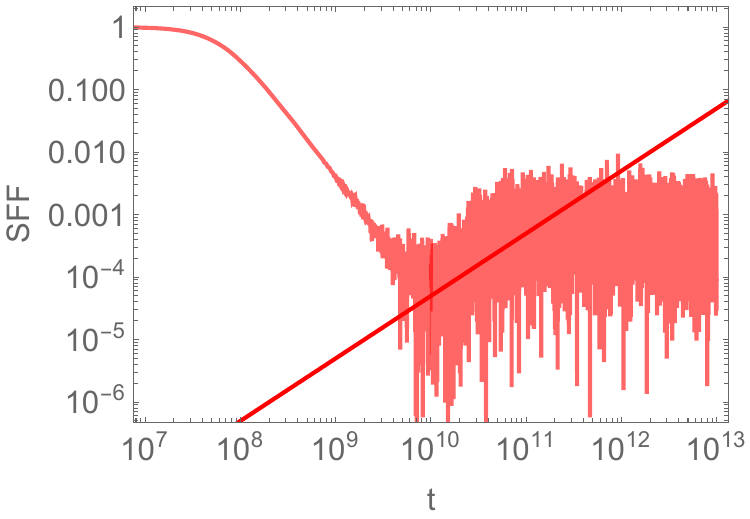} \label{}}
\quad
     \subfigure[$\sigma=0.0022$ (GUE)]
     {\includegraphics[width=4.4cm]{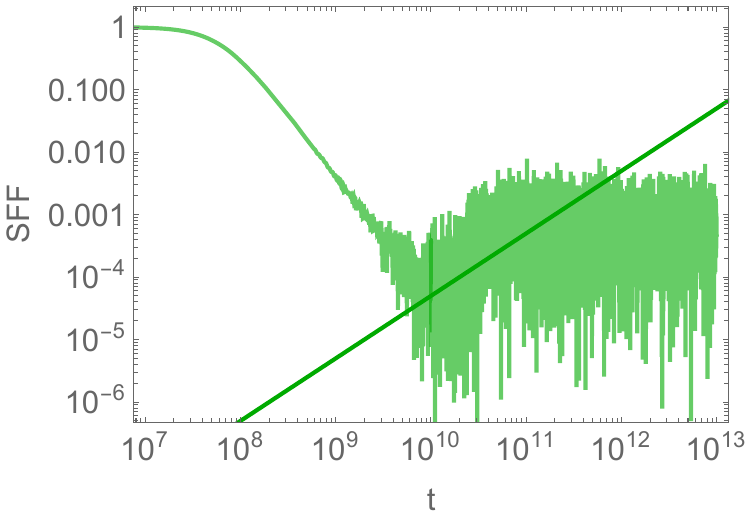} \label{}}
\quad
     \subfigure[$\sigma=0.0028$ (GOE)]
     {\includegraphics[width=4.4cm]{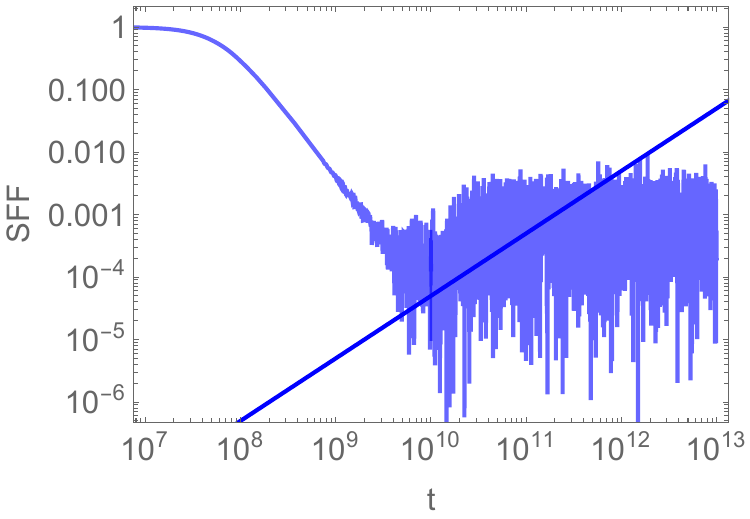} \label{}}
     
     \subfigure[$\sigma=0.0011$ (GSE)]
     {\includegraphics[width=4.4cm]{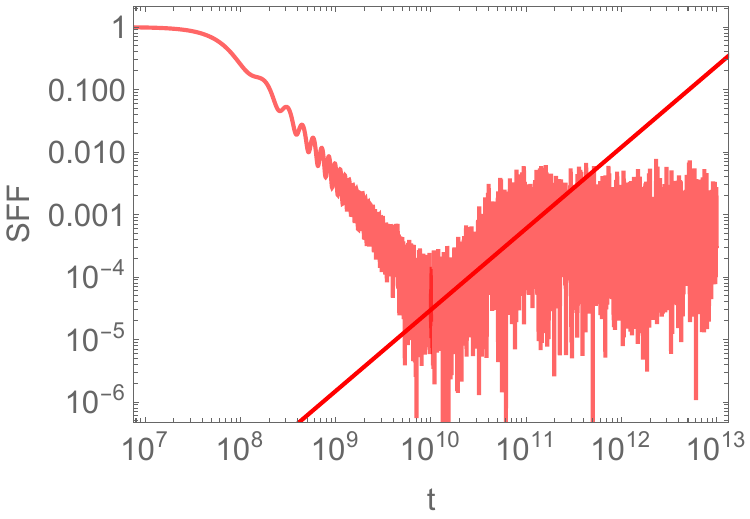} \label{}}
\quad
     \subfigure[$\sigma=0.0016$ (GUE)]
     {\includegraphics[width=4.4cm]{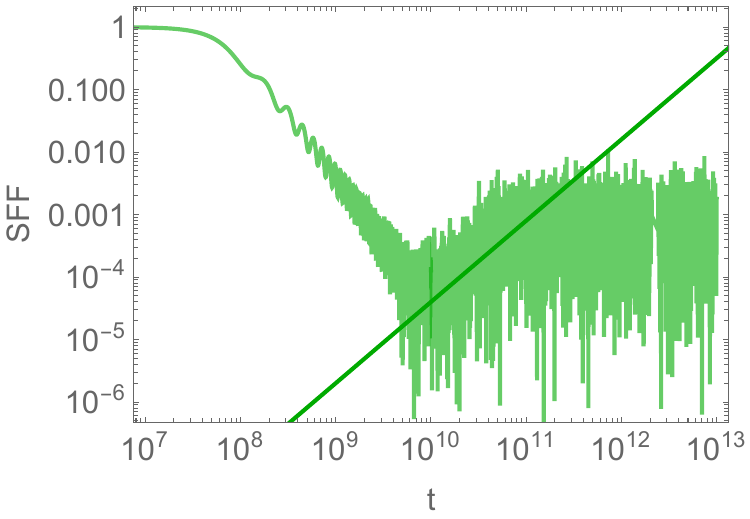} \label{}}
\quad
     \subfigure[$\sigma=0.0020$ (GOE)]
     {\includegraphics[width=4.4cm]{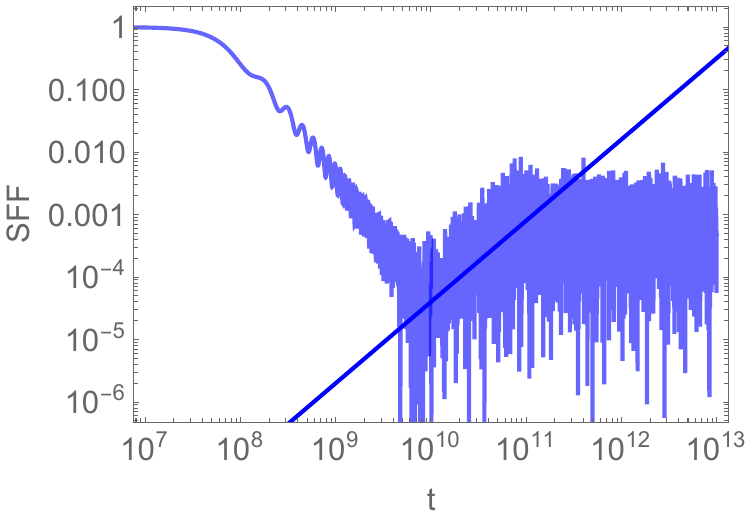} \label{}}
     
     \subfigure[$\sigma=0.0004$ (GSE)]
     {\includegraphics[width=4.4cm]{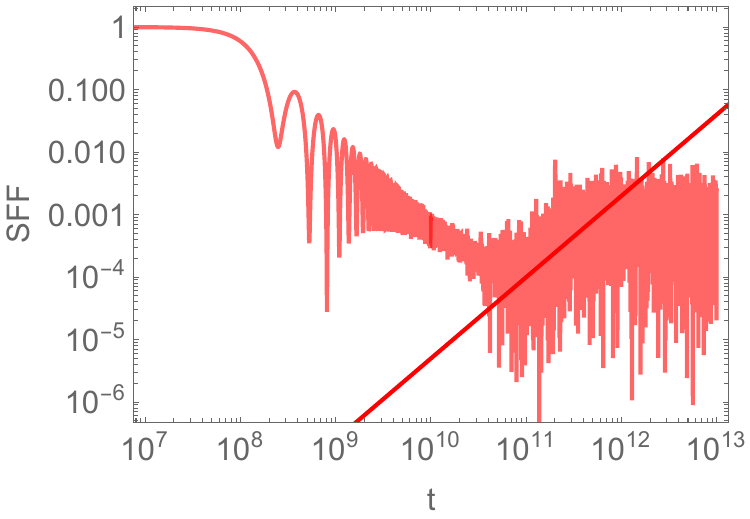} \label{}}
\quad
     \subfigure[$\sigma=0.0005$ (GUE)]
     {\includegraphics[width=4.4cm]{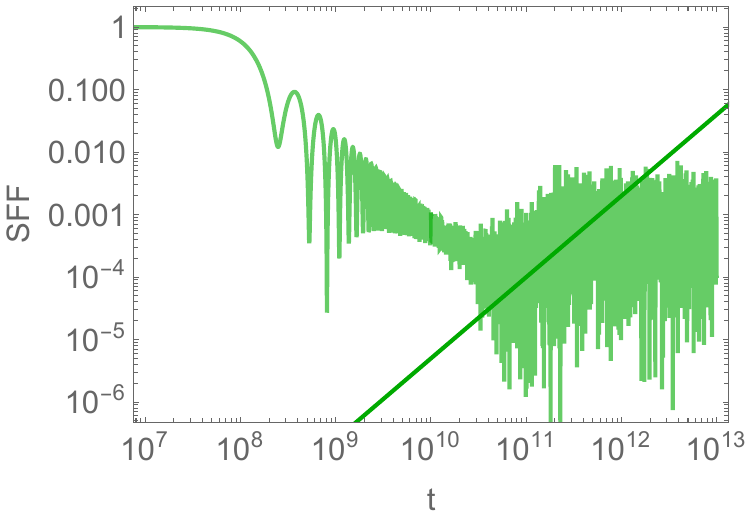} \label{}}
\quad
     \subfigure[$\sigma=0.0007$ (GOE)]
     {\includegraphics[width=4.4cm]{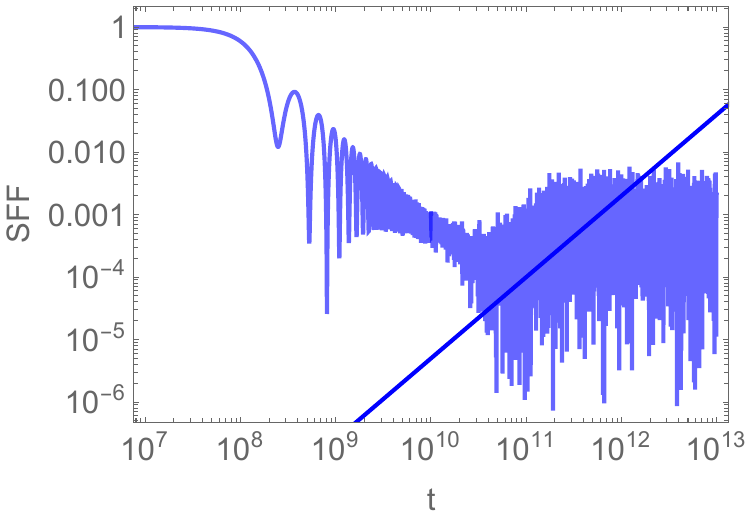} \label{}}
 \caption{Spectral form factor of scalar fields for $d=2$ (a-c), $d=100$ (d-f), and $d=500$ (g-i). The slope of the ramp is also displayed in a log-log plot for further clarity.}\label{DRAFTFIG7}
\end{figure}

\paragraph{Spectral form factor and the ramp structure.}
The SFF, defined in \eqref{SFFFORBW}, is shown in Fig. \ref{DRAFTFIG7} for scalar fields with the same parameters as in Fig. \ref{DRAFTFIG3}. The characteristic ramp structure appears, consistent with RMT predictions~\cite{Brezin:1997aa,Cotler:2016fpe}.
\begin{figure}[]
\centering
     \subfigure[$\sigma=0$]
     {\includegraphics[width=6.2cm]{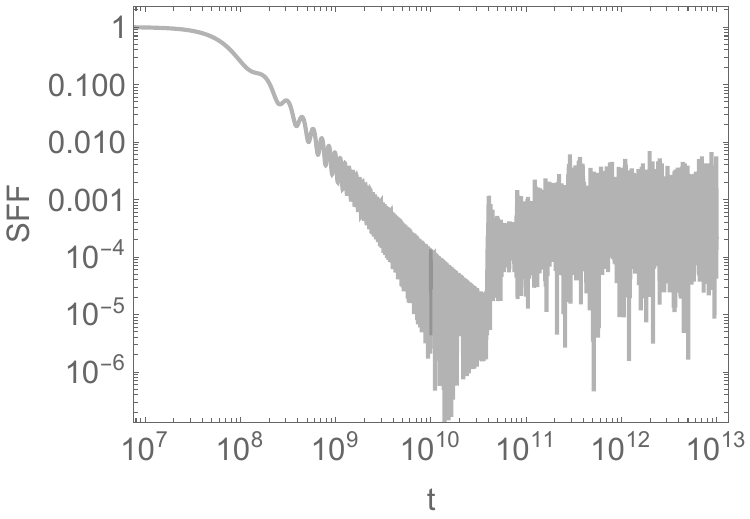} \label{}}
\quad
     \subfigure[$\sigma=0.2$]
     {\includegraphics[width=6.2cm]{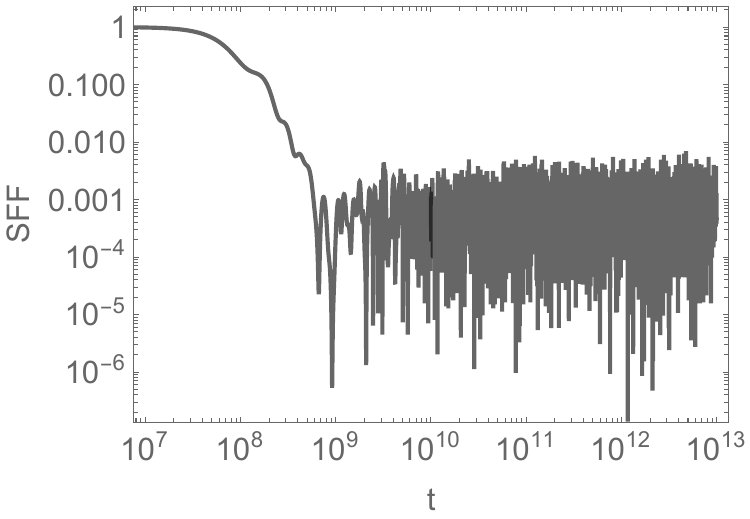} \label{}}
 \caption{Spectral form factor of scalar fields for $d=100$ with $\sigma=0$ (left) and $0.2$ (right).}\label{DRAFTFIG8}
\end{figure}

In Fig. \ref{DRAFTFIG8}, we also display the limiting cases $\sigma=0$ with $\sigma=0.2$: for $\sigma=0$, the SFF exhibits anomalous steep ramp, whereas for $\sigma=0.2$ (Poisson statistics) the ramp disappears. The onset time of the ramp depends on the spatial dimension $d$, shifting to later times as $d$ increases: e.g., compare Fig. \ref{DRAFTFIG7} (a), (d), and (g). This dimensional behavior parallels the feature of the Krylov complexity below.

\begin{figure}[]
\centering
     \subfigure[$d=2$]
     {\includegraphics[width=4.5cm]{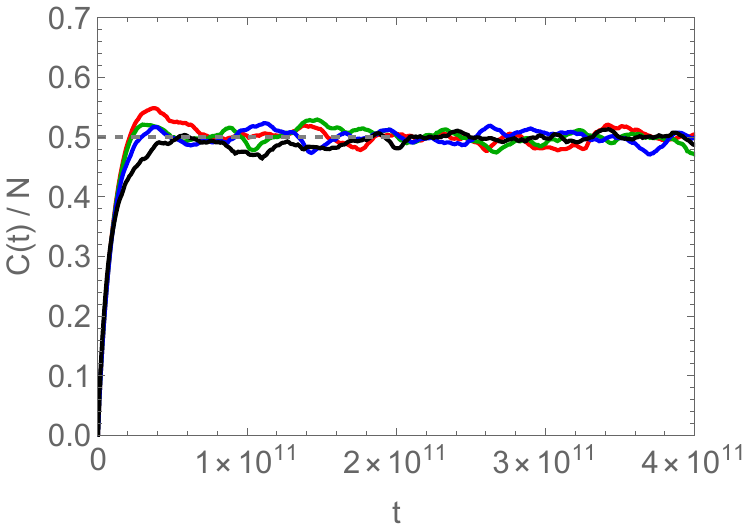} \label{}}
\quad
     \subfigure[$d=100$]
     {\includegraphics[width=4.5cm]{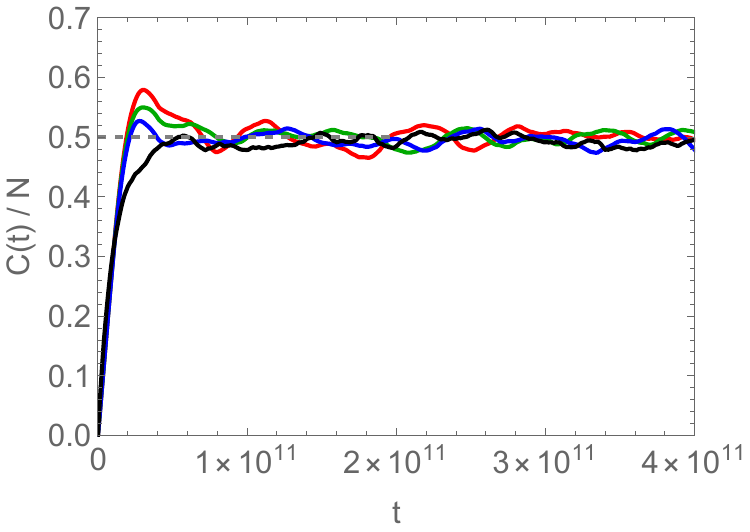} \label{}}
\quad
     \subfigure[$d=500$]
     {\includegraphics[width=4.5cm]{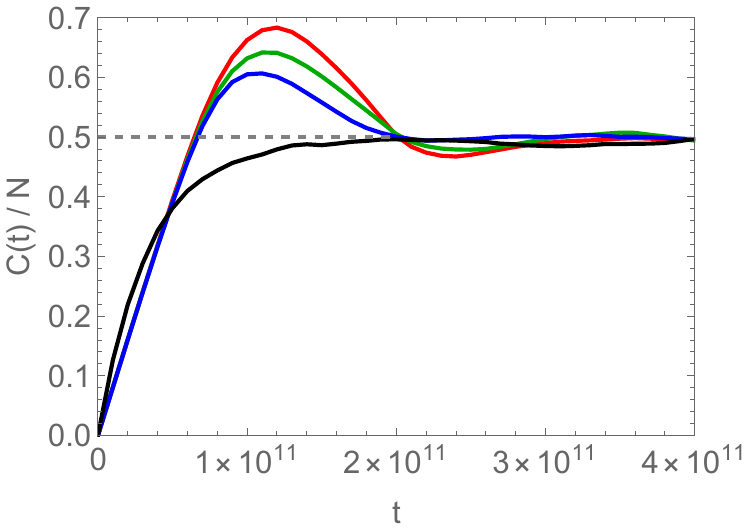} \label{}}
 \caption{Krylov complexity of scalar fields for $d=2$, $d=100$, and $d=500$. Parameters and color conventions match those used in the level-spacing distribution (Fig.~\ref{DRAFTFIG3}) and the spectral form factor (Fig.~\ref{DRAFTFIG7}). The emergence of the peak structure signals quantum chaotic behavior.}\label{DRAFTFIG9}
\end{figure}

\paragraph{Krylov complexity and the characteristic peak.}
We next examine the Krylov complexity, defined in \eqref{eq:Krylov complexity} and scaled by the system size $N$ (i.e., the total number of normal modes), so that $C(t)/N \to 1/2$ at late times \eqref{SFFCRE}: see Fig. \ref{DRAFTFIG9}. For $\sigma$ corresponding to GSE, GUE, and GOE, Krylov complexity develops a characteristic peak, consistent with earlier connections between Krylov complexity and RMT~\cite{Balasubramanian:2022tpr}. For Poisson statistics ($\sigma=0.2$), this peak vanishes.\footnote{We also find that the Lanczos coefficients exhibit qualitatively same behavior to those in BTZ cases~\cite{Jeong:2024jjn}.}

At $\sigma=0$, the Krylov complexity exhibits not only a pronounced peak but also strong late-time oscillations (omitted here for clarity in the chaos vs. Poisson comparison; see the grey data in Fig.~7 of \cite{Jeong:2024jjn} for details). This behavior mirrors the features of saddle-dominated scrambling observed in models such as the Lipkin-Meshkov-Glick model and inverted harmonic oscillators~\cite{Huh:2023jxt}. In this vanishing-variance case, three signatures emerge simultaneously: (i) delta-function-like level spacings, (ii) a steep SFF ramp, and (iii) a peaked, oscillatory Krylov complexity.\footnote{Comparable features appear in toy models with logarithmic spectra $E_n=\log n$~\cite{Das:2023yfj}.} These observations suggest that the $\sigma=0$ brickwall model in generic dimensions captures aspects of saddle-dominated scrambling, extending the argument first made for $d=2$ in \cite{Jeong:2024jjn}, though a definitive interpretation remains open.

We conclude this section with two observations on the dimensional dependence of Krylov complexity. First, as $d$ increases, the peak shifts to later times, consistent with the delayed onset of the SFF ramp. This correlation of timescales parallels findings in RMT~\cite{Erdmenger:2023wjg} and in spin-chain models~\cite{Camargo:2024deu}. Second, the peak amplitude shows a systematic growth with $d$, as illustrated in Fig.~\ref{DRAFTFIG9}(a-c). A plausible, albeit heuristic and non-rigorous, explanation attributes the effect to the enlarged phase-space volume of higher-dimensional systems: additional momentum directions at fixed energy increase the density of states, allowing more different $(p_x, p_y, \cdots)$ combinations to yield the same total energy. Consequently, the expanded phase space may supply extra degrees of freedom for the initial state to spread, thereby enhancing the peak in the Krylov complexity.

An analogous effect arises in random matrix ensembles, where GSE and GUE exhibit higher peaks than GOE. For example, as shown analytically in \cite{Huh:2024ytz} (cfr. eq.(10) vs. eq.(13)), within $2\times 2$ random matrix toy models, the GUE admits more configurations associated with a given eigenvalue than the GOE, leading to a correspondingly larger peak.

%
\section{Conclusions}\label{SEC4}
In this work, we have investigated the brickwall model~\cite{tHooft:1984kcu} as a framework for probing the quantum chaotic dynamics of hyperbolic AdS$_{d+1}$ black holes across generic dimensions. By quantizing probe scalar fields under Gaussian-distributed boundary conditions on the stretched horizon, we demonstrated that the model captures major features of quantum chaos, in alignment with the predictions of random matrix theory. Our analysis confirmed that while the logarithmic spectrum of normal modes dominates in the BTZ case $(d=2)$~\cite{Das:2022evy,Das:2023ulz,Jeong:2024jjn}, higher-dimensional black holes exhibit a transition toward power-law spectra. This dimensional sensitivity enriches our understanding of black hole microstates and extends the applicability of the brickwall model beyond low-dimensional settings.

Through diagnostics such as level spacing distributions, spectral form factors, and Krylov complexity, we established that the chaotic features persist robustly in moderate higher dimensions, yet gradually fade in the parametric large-$d$ limit. In this regime, degeneracy of normal modes emerges, effectively erasing chaotic signatures and reducing the system to a non-chaotic phase. This finding resonates with the broader large-$d$ perspective in gravitational physics~\cite{Emparan:2013moa,Emparan:2020inr}.

The results thus emphasize the utility of the brickwall model as a tractable, analytically controlled tool for connecting black hole microphysics with universal aspects of quantum chaos, while also highlighting its limitations in the asymptotic dimensional regime. Importantly, the requirement to retain angular momentum quantum numbers $J$ in the normal-mode analysis proved essential~\cite{Das:2022evy,Das:2023ulz,Jeong:2024jjn}: only with $J$-dependence do chaotic signatures manifest at least under Gaussian-distributed boundary conditions, underscoring that degrees of freedom beyond thermodynamic quantities are indispensable to the quantum chaotic description.

A natural next step is to explore whether and how the brickwall model can be adapted to two-dimensional black hole systems,\footnote{See \cite{Mann:1990fk,Kim:2007nh} for the early discussion on the brickwall model in two-dimensional spacetime.} notably Jackiw-Teitelboim gravity and sine-dilaton gravity which can be more appropriate to study the quantum mechanical holographic dual, such as SYK models.\footnote{In recent years, two-dimensional gravity has served as a useful setting for holographic duality. For instance, studies~\cite{Heller:2024ldz,Rabinovici:2023yex,Lin:2022rbf,Fu:2025kkh,Heller:2025ddj} show that Krylov complexity in the certain limit of SYK admits a gravity dual (geodesic lengths) in Jackiw-Teitelboim and sine-dilaton gravity.} In these cases, angular momentum labels $J$ are absent, raising the question of what degrees of freedom and which boundary conditions may play an analogous role in sustaining chaotic dynamics. Possible strategies include: (I) introducing effective boundary fluctuations or exotic matter profiles to mimic the role of $J$; (II) examining whether fluctuations in the dilaton or near-horizon degrees of freedom can seed the requisite spectral statistics; (III) embedding the two-dimensional models in higher-dimensional constructions where additional compact directions may supply effective angular labels.

Such generalizations could clarify whether the emergence of quantum chaos in black hole spectra is inherently tied to angular momentum structure or whether it can arise from more universal features of horizon dynamics. Extending the brickwall framework into these two-dimensional gravity settings may ultimately offer a bridge between holographic models of chaos and solvable lower-dimensional black hole theories, advancing our understanding of universality and microscopic structure in quantum gravity.

%
\acknowledgments
We would like to thank {Suman Das, Kyoung-Bum Huh, Juan F. Pedraza, Jos\'e Manuel Begines S\'anchez} for valuable discussions and correspondence. 
HSJ was supported by an appointment to the JRG Program at the APCTP through the Science and Technology Promotion Fund and Lottery Fund of the Korean Government. HSJ was also supported by the Korean Local Governments -- Gyeongsangbuk-do Province and Pohang City.
HSJ was also supported by the Spanish MINECO ‘Centro de Excelencia Severo Ochoa' program under grant SEV-2012-0249, the Comunidad de Madrid ‘Atracci\'on de Talento’ program (ATCAM) grant 2020-T1/TIC-20495, the Spanish Research Agency via grants CEX2020-001007-S and PID2021-123017NB-I00, funded by MCIN/AEI/10.13039/501100011033, and ERDF A way of making Europe.
This work was supported by the Basic Science Research Program through the National Research Foundation of Korea (NRF) funded by the Ministry of Science, ICT $\&$ Future Planning (NRF-2021R1A2C1006791) and the AI-based GIST Research Scientist Project grant funded by the GIST in 2025. This work was also supported by Creation of the Quantum Information Science R$\&$D Ecosystem (Grant No. 2022M3H3A106307411) through the National Research Foundation of Korea (NRF) funded by the Korean government (Ministry of
Science and ICT).
All authors contributed equally to this paper and should be considered as co-first authors.

%
\bibliographystyle{JHEP}
\providecommand{\href}[2]{#2}\begingroup\raggedright\endgroup

\end{document}